\documentclass{article}
\usepackage{spconf,amsmath,amsfonts,graphicx}
\usepackage[hyphens]{url}
\usepackage{hyperref}
\usepackage{todonotes}
\usepackage{pgfplots}
\usepackage{tikz}
\usetikzlibrary{shapes}
\usepackage{xcolor}
\usepackage{balance}
\usepackage[inline]{enumitem}

\definecolor{fc}{HTML}{C5E0DC}
\definecolor{cost}{HTML}{F1D4AF}
\tikzset{cost/.style={black,
draw=black,
fill=cost,
rectangle,
rounded corners=10pt,
inner sep=3pt,
minimum height=0.8cm}}
\tikzset{net/.style={black,
draw=black,
fill=fc,
rectangle,
minimum height=1cm}}

\pgfplotsset{my style/.append style={axis x line=middle,axis y line=middle}}
\pgfplotsset{compat=1.14}

\title{SEMI-SUPERVISED AND POPULATION BASED TRAINING FOR VOICE COMMANDS RECOGNITION}
\name{Oguz H. Elibol, Gokce Keskin, Anil Thomas
}
\address{
Intel AI Lab, Santa Clara, CA
}

\begin{document}
% \ninept
%
\maketitle
\begin{abstract}
We present a rapid design methodology that combines automated hyper-parameter tuning with semi-supervised training to build highly accurate and robust models for voice commands classification. Proposed approach allows quick evaluation of network architectures to  fit performance and power constraints of available hardware, while ensuring good hyper-parameter choices for each network in real-world scenarios. Leveraging the vast amount of unlabeled data with a student/teacher based semi-supervised method, classification accuracy is improved from 84\% to 94\% in the validation set. For model optimization, we explore the hyper-parameter space through population based training and obtain an optimized model in the same time frame as it takes to train a single model.
\end{abstract}
\section{Introduction}
\label{sec:intro}

Recognition of short speech utterances, keywords, or voice commands as will be referred to in this paper is of interest in multiple domains. We already are experiencing an increase of speech interactions with devices around us through voice user interfaces \cite{Porcheron2018-rm}. However, high power and processing requirements for accurate and robust detection of voice commands can limit human interaction. Thus, it is of great interest to be able to rapidly design models that can fit within a given hardware with best possible performance. Re-purposing a deep neural net model for new hardware constraints usually require re-tuning of hyper-parameters and re-training which can be very time consuming. Besides hardware constraints for such applications, there is still the typical problem of not having enough labeled data to realize a robust system. Even in the presence of vast amount of labeled data, the environments in which these systems are deployed may have a different distribution of inputs compared to the training data, leading to significant decrease in accuracy and robustness. 

In this work we tackle these problems through a combination of techniques. First, we show that robust and accurate voice recognition systems can be trained through a semi-supervised technique which has been shown to be very effective in the vision domain\cite{Tarvainen2017-fg}. This approach also allows for systems to continually learn after deployment from unlabeled data, although this aspect of semi-supervised training is beyond the scope of this current study, and is a part of our future work. Second, automated hyper-parameter tuning of model parameters with Population Based Training \cite{Jaderberg2017-ed} can match or exceed the validation accuracy of hand-tuned models.  This automated approach helps rapid customization of model size to fit a specific hardware constraint while ensuring high validation accuracy. 

We demonstrate the effectiveness of our approach on a classification task using the Speech Commands Dataset \cite{Warden2018-bh}. This dataset contains one-second long utterances for a set of 30 words, with varying quality and different accents.  Models are evaluated according to the classification accuracy in the validation set, defined as correctly labeling an input sample into one of twelve possible classes (yes, no, up, down, left, right, stop, go, on, off, unknown word, silence). The amount of unlabeled data ($>$150k utterances) is greater compared to the labeled data (about 65k) in the dataset, and the unlabeled data contains words that have not been included in the labeled dataset, making it realistic for deployment. 

In Section \ref{sec:related_work} we review the related work. In Section \ref{sec:student_teacher} we present our data processing, model selection, and semi-supervised training methodology for training our model and present our results. In Section \ref{sec:pbt} we present our methodology for hyper-parameter tuning and model adaptations based on population based training and also present the results. In Section \ref{sec:discussion} we conclude with a discussion of the strengths and weaknesses of the techniques used, and possible future directions for this work. 

\section{Related Work}
\label{sec:related_work}

Usually to improve the accuracy and robustness of a model an ensemble of models are used. In one such example\cite{Solovyev2018-zy}, authors used an ensemble of many strong neural network models (VGG16, Xception, ResNet50, InceptionV2, InceptionV3) in order to achieve a high performing model on the Speech Commands Dataset \cite{Warden2018-bh}. However, deployment of such a model with varying hardware requirements is unrealistic from both an inference and training perspective for practical applications. 

Our work on the semi-supervised part is closely related to~\cite{Tarvainen2017-fg}, where the authors show that using a combination of noise augmentation and student-teacher model results in very sample efficient models for vision workloads. The work on population based training for hyper-parameter tuning and model adaptation is inspired by the work done by \cite{Jaderberg2017-ed}, in which they show that evolutionary techniques can beat hand tuning of deep neural net models and the time taken for the  technique is not any longer than training a single model given the computational resources.

\section{Semi-Supervised Deep Neural Net Training for Voice Commands}
\label{sec:student_teacher}

When training a deep learning model for speech, there are multiple options for signal representation. Using spectrograms have been shown to be effective features in speech processing \cite{Deng2013-au}. Thus after some experimentation, we also chose to pre-process the raw audio signal using mel scaled spectrograms. Comparing this route to using the raw data (model resulted in being too complex) or using MFCC features (saw a degradation of performance) as the input to the model, we saw an increase in final performance.

Following content-preserving data augmentation techniques are used to increase model robustness: 
\begin {enumerate*}[label=\itshape\alph*\upshape)]
\item Noise superposition (0.1, 0.15, 0.20, 0.25)
\item Time stretch (0.81, 0.93, 1.07, 1.23),
\item Pitch shift (-3.5, -2.5, -2.0, -1.0, 1.0, 2.0, 2.5, 3.5).
\end {enumerate*} Both the background noise sounds provided with the dataset and generated random Gaussian noise are used as noise sources. 

\begin{figure}[t]
    \centering
    \includegraphics[width = 8 cm]{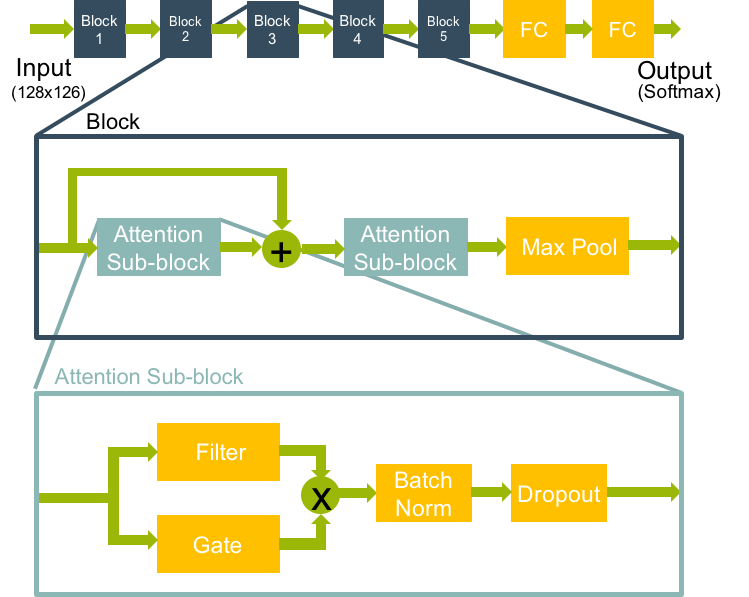}
    \caption{
    	Proposed network has five identical blocks (top of the figure) followed by two fully connected (FC) layers. \textbf{Block}: Consists of two attention sub-blocks and a max pooling layer to reduce the height and width of the feature maps by 2. \textbf{Attention Sub-block}: \textbf{Filter} is a convolutional filter with a ReLU output, whereas  \textbf{Gate} is also a convolutional filter of the same size but with a sigmoid output. The output of the \textbf{Filter} and the \textbf{Gate} are element-wise multiplied. Because the output of the \textbf{Gate} is between 0 and 1, this could be thought as gating (attending to, or ignoring) certain portions of the \textbf{Filter} output. Gated output is followed by batch norm and dropout. 
    }
    \label{fig:network}
\end{figure}

Due to the temporal brevity of our signal we chose to base our model on a 2D convolutional network. Our general approach in finding the final network topology was to start with simple networks and add network capacity until observing diminishing returns on validation accuracy. We also experimented with and incorporated several popular network motifs present in literature based on the measured performance gains on our dataset. One of these is using residual connections in the network \cite{He2015-rh}, and as expected, we observed the ability to increase the depth of our network to improve performance by enabling the flow of the gradient signal from the output. Second, we used an attention module in our convolutional network (similar to \cite{Van_den_Oord2016-sb}) in order to attend to specific components of the output feature maps at each stage. The resulting architecture used for this work is shown in Figure \ref{fig:network}.

We used the Speech Commands Dataset \cite{Warden2018-bh} for training our network. Using purely supervised training the top model validation accuracy was 84\%. We were able to improve the performance to 94\% through a variant of the semi-supervised technique that was previously shown to work well for vision applications \cite{Tarvainen2017-fg}. Briefly, the technique used is a combination of the two different ways of extracting information from unlabeled data: 
\begin {enumerate*}[label=\itshape\alph*\upshape)]
\item If two different noisy version of the same signal are introduced to the model, the model is supposed to give the same answer. Hence, the distance of the model output distributions between these two noisy inputs can be used as a signal to train the model
\item Assuming a teacher model that performs slightly better than the model being trained is available, it is possible to use the output of the teacher model as a label to training (student) model in the absence of labels. The teacher model is realized by the exponential moving average of the parameters (weights) of the model being trained (student). The basic intuition is that this teacher network is formed by ensembling the trained network in time, hence performs marginally better than the model (student) that is being trained.
\end {enumerate*}

\begin{figure}[t]
    \centering
    \includegraphics[width = 8 cm]{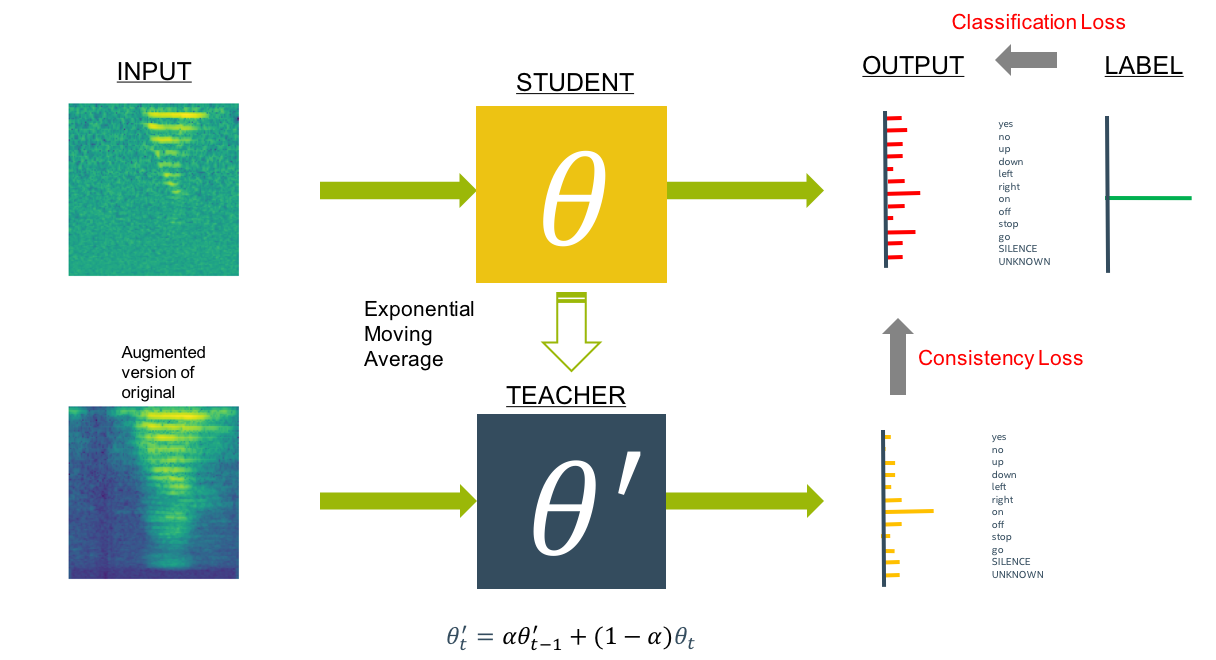}
    \caption{
    	Scheme for semi-supervised training. The network being trained (student) is represented as $\theta$, and the teacher network is the exponential moving average of $\theta$, represented as $\theta'$. A differently augmented version of the same signal is fed into both the student and teacher networks. 
    }
    \label{fig:StudentTeacher}
\end{figure}

Figure \ref{fig:StudentTeacher} shows our training procedure. A minibatch of data is introduced with mixed labeled and unlabeled data. For samples that lack the labels the distance (KL divergence) between the output distributions of the teacher network and the trained network is used as a loss signal. This is called the consistency loss, enforcing consistency between the student and the teacher.  For data that is labeled a cross entropy loss between the model output and the label (classification loss) is used in addition to the consistency loss. Because the total loss is a linear combination of these two losses, the weighting of these losses is another hyper-parameter for the model (which is tuned using population based training presented in Section \ref{sec:pbt}). 

\begin{figure}[t]
    \centering
    \includegraphics[width = 8 cm]{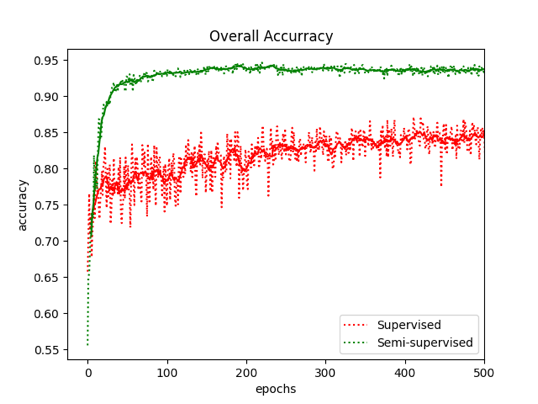}
    %\caption{Validation accuracy as a function of training epoch number shown for supervised (red curve) and semi-supervised (green curve) training.}
    \label{fig:accuracy}
\end{figure}

\begin{figure}[t]
    \centering
    \includegraphics[width = 7.5 cm]{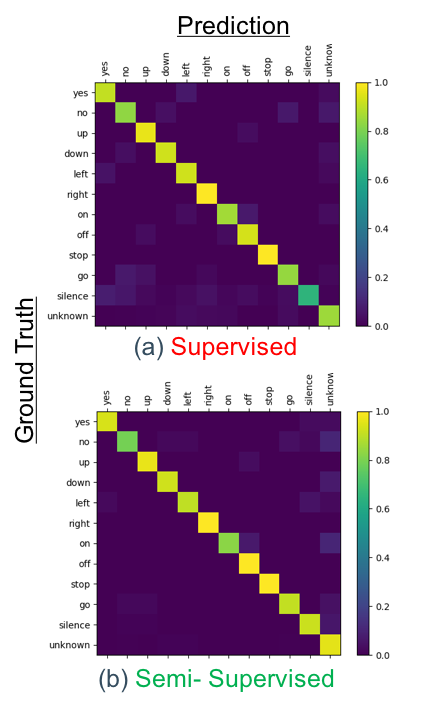}
    \caption{Confusion plots for supervised and semi-supervised training of the same model. The top plot shows the results of supervised technique and the bottom one shows the semi-supervised technique.
    }
    \label{fig:confusion}
\end{figure}

The validation accuracy as a function of training epochs of the same network being trained using supervised versus semi-supervised technique is shown in Figure \ref{fig:accuracy}. Using semi-supervised training, we both get higher final accuracy, and also faster convergence to top accuracy. The main failure mode of the supervised model is the confusion of the speech commands with silence and unknown words (Figure \ref{fig:confusion}). This is not surprising given that the model never sees any of the words that are not labeled which are in the test set. This problem is mostly alleviated in the semi-supervised approach as the model is trained on a much more diverse set of commands owing to the ability of being able use the unlabeled data.

\section{Population based training for hyper-parameter tuning and model adaptation}
\label{sec:pbt}

We have used population based training for tuning some of the hyper-parameters for our model. Briefly, 50 models are launched to train in parallel, each having their hyper-parameters of interest initialized randomly in a predetermined range. In this specific study we aimed to tune three critical hyper-parameters - dropout, consistency weight and the learning rate of the model.

At the end of every 2 epochs of training each of the models that are being trained in parallel are evaluated on a holdout data set (not the final validation set), and the best 10\% of the population (5 models) are selected. Another 50 models are launched to trained in parallel, but initialized with the weights of these chosen individuals, with the hyper-parameter being tuned slightly perturbed for this population. This is very similar to mutating the parameters as done in genetic algorithms, but in this case the perturbation is directly on the hyper-parameter (not on vectorized representations) and there is no cross-over being used. Because each model is concurrently trained, within the time it takes to train a single model we can have a model that is nearly optimally trained - although at the expense of computational resources. Another advantage of using this method is that hyper-parameters are dynamically adjusted, because a new value can be assigned every 2 epochs. Hence, there is no need for experimenting with a hand designed hyper-parameter schedule for these parameters. 

\begin{figure}[t]
\begin{minipage}[b]{1.0\linewidth}
  \includegraphics[width = 7.5 cm]{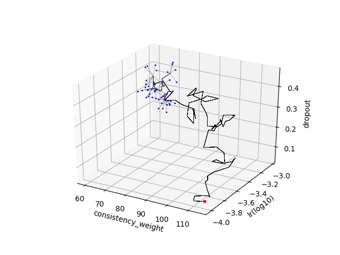}
  \centerline{(a) Hyper-parameter trajectory (schedule)}\medskip
\end{minipage}
\begin{minipage}[b]{1.0\linewidth}
  \centering
  \includegraphics[width = 6 cm]{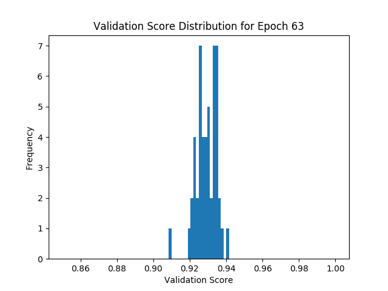}
  \centerline{(b) Final distribution of population}\medskip
\end{minipage}
\caption{(a)Visualization of the trajectory of the population in the hyper-parameter space. Red dot is the starting point, black lines represents the trajectory taken through training and the blue dots represent the final locations of the current population. (b) Final distribution of the validation scores for the trained population.}
\label{fig:pbt}
\end{figure}

The trajectory of the population in the hyper-parameter space can be visualized to gain intuition into the process as shown in Figure \ref{fig:pbt} (a). Interestingly, best schedule found for the hyper-parameters seems to have a specific direction instead of a fully random walk, and studying this behavior further can provide additional insight into the training process. We have used this technique to ensure a good set of hyper-parameters and associated schedules, and studying the specific behavior was not pursued further for this study, but is an interest for future work. 

Figure \ref{fig:pbt} (b) shows the distribution of validation scores for the final population, and the best individual in the population surpasses 94\% final accuracy, which is also used as the final model.

Besides using population based training for rapidly tuning a given model as done in this case, it could also be used for the purposes of re-purposing the model for a different device. Usually this is done through compression techniques \cite{Cheng2018-to}. However when the size of the network has to be downsized dramatically, these techniques may not be enough. We used population based training to reduce the size of the model (250 Mb) down  to (1.5 Mb) to fit on an edge device, with a performance hit of only 4\% (down to 90\% from 94\%). Specifically a relatively shallow convolutional network was trained by the same methodology highlighted above to achieve this result.  

\section{Discussion}
\label{sec:discussion}

We presented a deep learning model and associated methods to alleviate challenges in training and deploying systems for speech commands recognition. Specifically we showed that using semi-supervised training significantly improves the accuracy and robustness of the model compared to using only supervised training, resulting in an increase of accuracy from 84\% to 94\% on the Speech Commands Dataset. Using the semi-supervised method outlined here also allows the possibility of learning after deployment for added robustness by periodic re-training on the (unlabeled) data observed during deployment. Although demonstrating this capability is out of the scope of our current work, it is a subject of our future work.

Further, we showed that population based training can be an effective tool for rapidly iterating through different models as it allows hyper-parameter tuning and schedule finding with no extra training time, which in turn allows to investigate and try out different network topologies rapidly without having to worry about tuning. One such advantage is to be able to change the model to fit the constraints of a given hardware for the specific application. Using population based training comes with significant demands for computational resources, as it is assumed that each individual in the population can be trained in parallel. Also, the technique only allows for search in the hyper-parameter space and not in the architecture space. The methodology for mutating the current parameter (picking the next set of parameters) can be improved by accounting for the past history of the individual to be mutated. Thus, despite the improvements in tackling the challenges in deployment and performance for voice commands recognition presented here, there is still room for improvement and these are subjects of our future work.

\newpage
\balance
\bibliographystyle{myieee}
{\ninept \bibliography{ref_gokce}}

\end{document}